\documentclass[conference,a4paper]{APSIPA2018}
\usepackage{multirow}
\usepackage{amsmath}
\usepackage[psamsfonts]{amssymb}
\usepackage{amsxtra}
\usepackage{threeparttable}
\usepackage{tabularx}
\usepackage[numbers,sort&compress]{natbib}
\usepackage{graphicx,accents}

\usepackage{hyperref}

\makeatletter
\DeclareRobustCommand{\cev}[1]{%
	\mathpalette\do@cev{#1}%
}
\newcommand{\do@cev}[2]{%
	\fix@cev{#1}{+}%
	\reflectbox{$\m@th#1\vec{\reflectbox{$\fix@cev{#1}{-}\m@th#1#2\fix@cev{#1}{+}$}}$}%
	\fix@cev{#1}{-}%
}
\newcommand{\fix@cev}[2]{%
	\ifx#1\displaystyle
	\mkern#23mu
	\else
	\ifx#1\textstyle
	\mkern#23mu
	\else
	\ifx#1\scriptstyle
	\mkern#22mu
	\else
	\mkern#22mu
	\fi
	\fi
	\fi
}

\makeatother

\begin{document}

\title{Error Reduction Network for DBLSTM-based Voice Conversion}

\author{%
\authorblockN{%
Mingyang Zhang\authorrefmark{1}\authorrefmark{2}, 
Berrak Sisman\authorrefmark{2}, 
Sai Sirisha Rallabandi\authorrefmark{2},
Haizhou Li\authorrefmark{2},
Li Zhao\authorrefmark{1}
}
\authorblockA{%
\authorrefmark{1}
 Key Laboratory of Underwater Acoustic Signal Processing of Ministry of Education, Southeast University, Nanjing, China\\
E-mail: zhangmy@seu.edu.cn, zhaoli@seu.edu.cn}
\authorblockA{%
\authorrefmark{2}
National University of Singapore, Singapore \\
E-mail: berraksisman@u.nus.edu, siri.gene@gmail.com, haizhou.li@nus.edu.sg}
}

\maketitle
\thispagestyle{empty}

\begin{abstract}
So far, many of the deep learning approaches for voice conversion produce good quality speech by using a large amount of training data. This paper presents a Deep Bidirectional Long Short-Term Memory (DBLSTM) based voice conversion framework that can work with a limited amount of training data. We propose to implement a DBLSTM based average model that is trained with data from many speakers. Then, we propose to perform adaptation with a limited amount of target data. Last but not least, we propose an error reduction network that can improve the voice conversion quality even further. The proposed framework is motivated by three observations. Firstly, DBLSTM can achieve a remarkable voice conversion by considering the long-term dependencies of the speech utterance. Secondly, DBLSTM based average model can be easily adapted with a small amount of data, to achieve a speech that sounds closer to the target. Thirdly, an error reduction network can be trained with a small amount of training data, and can improve the conversion quality effectively. The experiments show that the proposed voice conversion framework is flexible to work with limited training data and outperforms the traditional frameworks in both objective and subjective evaluations.
\end{abstract}

\section{Introduction}
Voice Conversion (VC) is a technology that modifies the speech of the source speaker to make it sounds like the target speaker. The voice conversion technology has been applied to many tasks, such as Text-to-Speech (TTS) system \cite{vc_tts}, speech enhancement \cite{TodaEnhence} and speaking assistance \cite{NAKAMURA2012134}. 

Voice conversion can be formulated as a regression problem of estimating a mapping function between the source and target features. Many state-of-the-art approaches for voice conversion are including Gaussian Mixed Model (GMM) \cite{Toda2007, takamichi2015modulation, tanaka2017speaker} which is based on the maximum-likelihood estimation of spectral parameter trajectory. Dynamic Kernel Partial Least Squares (DKPLS) \cite{Helander2012} integrates a kernel transformation into partial least squares to model nonlinearities as well as to capture the dynamics in the data. Sparse representation \cite{Wu2014, takashima2013exemplar, ccicsman2017sparse, berrak2} can be seen as a data-driven, non-parametric approach as an alternative to the traditional parametric approaches to voice conversion. Frequency warping based approaches \cite{tian2014correlation, Tian2015, Tian2016} aim to modify the frequency axis of source spectra towards that of the target. There are also some post-filter approaches for voice conversion to improve the speech quality \cite{7393764, 7760371}.

Recently, deep learning approaches became popular in the field of voice conversion. For example, Deep Neural Network (DNN) based approaches \cite{Chen2014, Nakashika2013, mohammadi2014voice} focus on spectrum conversion under the constraint of parallel training data and achieve high-quality speech by using a large amount of parallel training data. In addition, there have been some researches on variational autoencoder \cite{hsu2016voice, hsu2017voice} that effectively improve the conversion performance.

The above-mentioned voice conversion frameworks consider the frame features as individual components, and do not concern about the long-term dependencies of the speech sequences. Standard Recurrent Neural Networks (RNNs) can be used to solve this problem \cite{Nakashika2014, nakashika2015voice}, but it has limited ability in modeling context because of the vanishing gradient problem \cite{bengio1994learning}. Moreover, the standard RNNs can only capture the information from the former sequences and not the latter sequences.

To alleviate these problems, Deep Bidirectional Long Short-Term Memory (DBLSTM) has been proposed to perform voice conversion \cite{hochreiter1997long, graves2005framewise, Sun2015}, and achieves remarkable performance over the traditional DNN-based voice conversion framework \cite{Sun2015}. Moreover, DBLSTM has been successfully used in various tasks in the field of speech and language processing, such as Automatic Speech Recognition (ASR) \cite{graves2013hybrid, wollmer2013feature, zeyer2017comprehensive}, speech synthesis \cite{fan2014tts} and emotion recognition \cite{wollmer2010context, he2015multimodal}. 

Although the DBLSTM and DNN based voice conversion frameworks can achieve good voice conversion performance, they still suffer from the dependency of a large amount of training data which is not practical in real life. The remaining issue is to find a way to make a good use of limited data. Different from the previous studies, in this paper, we take advantage of the powerful deep learning framework DBLSTM, and propose a voice conversion framework that can produce high-quality speech under the constraint of limited parallel data. Specifically, we make the following contributions: 1) due to DBLSTM can achieve a remarkable voice conversion by considering the long-term dependencies of the speech utterance, we build a DBLSTM-based average model by using data from many speakers; 2) since the DBLSTM-based average model can be easily adapted with a small amount of data, we perform adaptation to the DBLSTM-based average model by using limited target data, to achieve a converted sound that is more similar to that of target; 3) an error reduction network can be trained with a small amount of training data from source and target, so we propose an error reduction network for the adapted DBLSTM framework, that can improve the voice conversion quality even further. Overall, we propose a DBLSTM-based voice conversion framework that can produce high-quality speech with a small amount of training data.

The rest of this paper is organized as follows: Section II explains the traditional DBLSTM-based approach of voice conversion. Section III describes our proposed voice conversion framework that is based on an error reduction network for the adapted DBLSTM-based approach. We report the objective and subjective results in Section IV. Section V concludes this paper.
\section{DBLSTM-based Voice Conversion}
The network architecture of BLSTM-based Voice Conversion is a combination of bidirectional RNNs and LSTM memory block, which can learn long-range contextual in both forward and backward directions. By stacking multiple hidden layers, a deep network architecture is built to capture the high-level representation of voice features. For bidirectional RNNs, the iteration functions for the forward sequence $\vec{h}$ and backward sequence $\cev{h}$ are as follows:
\begin{align}
\vec{h}_{t}&=\mathcal{H}(W_{x\vec{h}}x_{t}+W_{\vec{h}\vec{h}}\vec{h}_{t-1}+b_{\vec{h}})\\
\cev{h}_{t}&=\mathcal{H}(W_{x\cev{h}}x_{t}+W_{\cev{h}\cev{h}}\cev{h}_{t+1}+b_{\cev{h}})\\
y_{t}&=W_{\vec{h}y}\vec{h}_{t}+W_{\cev{h}y}\cev{h}_{t}+b_{y}
\end{align}
where $x,y,h,t$ are the input, output, hidden state and time sequence
respectively. A LSTM network consists of recurrently connected blocks, known as memory block. Every memory block contains self-connected memory cells and three adaptive and multiplicative gate units. The recurrent hidden layer function $\mathcal{H}$ of the LSTM network is implemented according to the following equations:
\begin{align}
i_{t}&=\sigma(W_{xi}x_{t}+W_{hi}h_{t-1}+W_{ci}c_{t-1}+b_{i})\\
f_{t}&=\sigma(W_{xf}x_{t}+W_{hf}h_{t-1}+W_{ci}c_{t-1}+b_{f})\\
c_{t}&=f_{t}c_{t-1}+i_{t}\tanh(W_{xc}x_{t}+W_{hc}h_{t-1}+b_{c})\\
o_{t}&=\sigma(W_{xo}x_{t}+W_{ho}h_{t-1}+W_{co}c_{t}+b_{o})\\
h_{t}&=o_{t}\tanh(c_{t})
\end{align}
where $i,f,o,c$ refer to the input gate, forget gate, output gate and the element of cells $C$ respectively. $\sigma$ is the logistic sigmoid function.

The overall framework of a DBSLTM-based voice conversion is shown in Fig. 1. In this framework, the three feature streams including the spectrum feature, $\log(F_{0})$ and the aperiodic component are converted separately. The spectrum feature is converted by the DBLSTM model. $\log(F_{0})$ is converted by equalizing the mean and the standard deviation of the source and target speech. And the aperiodic component is directly copied to synthesize the converted speech. The whole utterance is treated as input so that the system can access the long-range context in both forward and backward directions. In this paper, we propose to use DBLSTM to perform voice conversion under limited training data.

\begin{figure}[t]
	\begin{center}
		\includegraphics[width=80mm]{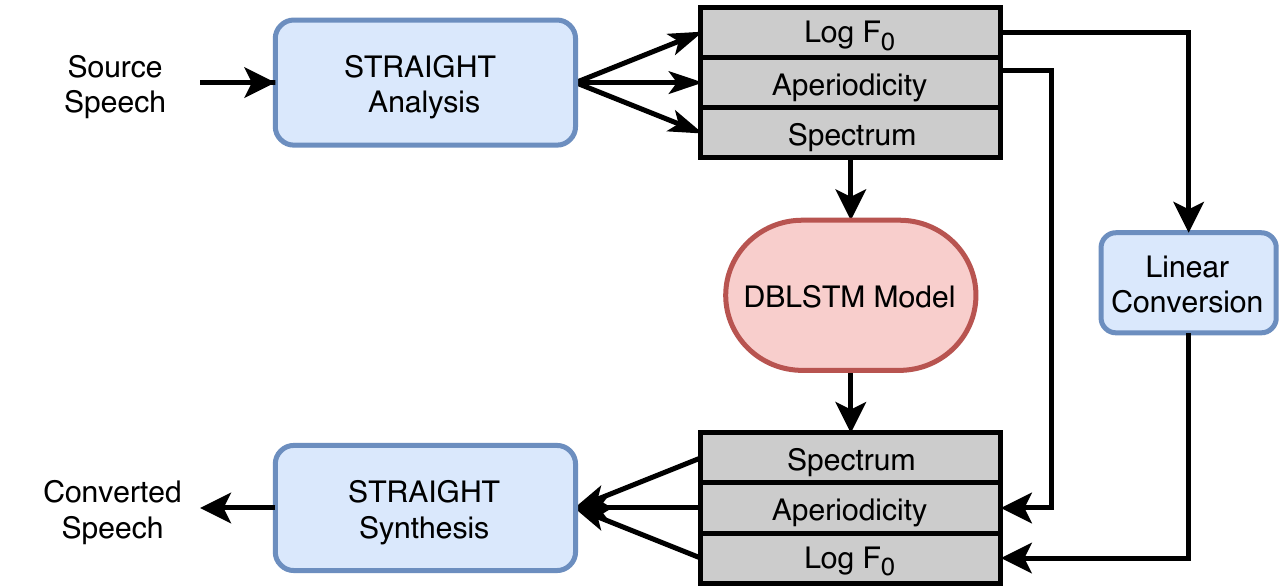}
	\end{center}
	\caption{DBLSTM-based voice conversion framework.}
	\vspace*{7pt}
\end{figure}

\section{Error Reduction Network for Adapted DBLSTM-based Voice Conversion}
The DBLSTM-based voice conversion has a good performance while it needs a large amount of parallel data from source speaker and target speaker, which is expensive to collect in practice. To solve this problem, we propose an error reduction network for adapted DBLSTM-based approach.

\begin{figure*}[ht]
	\begin{center}
		\includegraphics[width=160mm]{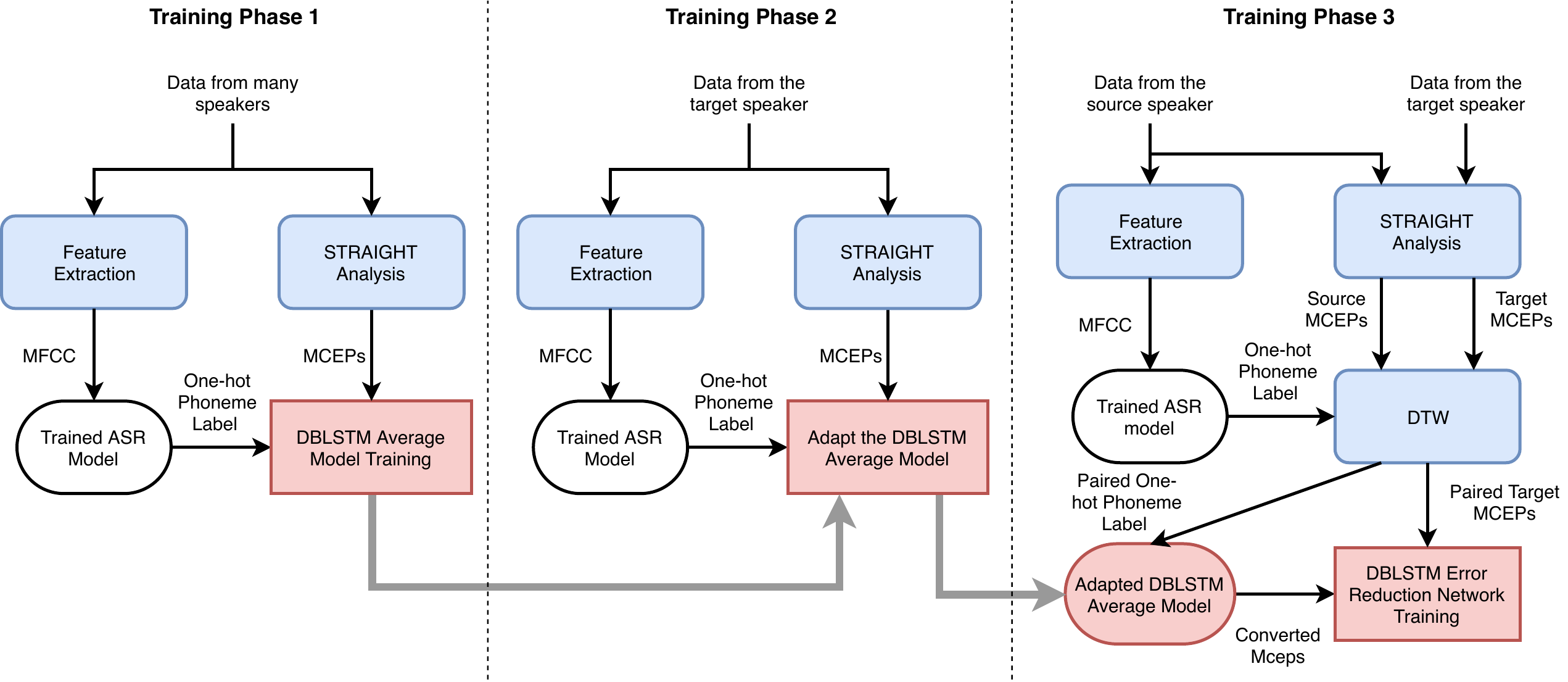}
	\end{center}
	\caption{The proposed DBLSTM based VC framework. In training phase 1, we exclude the data from the source and the target speakers. In training phase 2, we only use the data from the target speaker. In training phase 3, we use the same sentences from both source and target speakers.}
	\vspace*{7pt}
\end{figure*}

\subsection{Training Phase}

As illustrated in Fig. 2, the proposed approach can be divided into three training phases. In training phase 1, an average DBLSTM model is trained for the one-hot phoneme label to Mel-cepstral coefficients (MCEPs) mapping, using the data from many speakers except the source speaker and the target speaker. MCEPs are the Mel Log Spectral Approximation (MLSA) parameters which approximate Mel-Frequency Cepstral Coefficients (MFCCs). A trained ASR system is used to extract the phoneme information of the input speech. The input of the ASR model is MFCC feature of the speech frame. The output is a one-hot phoneme label vector that indicates the phoneme information of the speech frame. Then a DBLSTM-based model is trained to get the mapping relationship between the one-hot phoneme label and the corresponding MCEPs which are extracted by STRAIGHT \cite{straight}. We call this framework as Average Model, it can only generate MCEPs of an average voice of the speakers whose data are used.

In training phase 2, the average model is adapted using a small amount of data from the target speaker. The adaptation is similar to the training of the average model, except the initialized network is the trained average model and the training data is the target speech. After the adaptation, the output of the adapted model will be closer to the target speaker. We call this framework as adapted average model. However, there always exists an error between the converted features and the target features. This error degrades both quality and similarity of the converted speech \cite{wu2017denoising}. To reduce such error, we propose an error reduction network after the adapted average model.

Training phase 3 involves the error reduction network which is essentially an additional DBLSTM network, used to map the converted MCEPs to the target MCEPs. The error reduction network brings the final output
MCEPs features closer to the target speaker. The same utterances have been used in the adapted average model of the target speaker and the parallel data of the source speaker are used in the error reduction network. The same ASR system is used to generate the one-hot phoneme label of the target speech. MCEPs features from the same sentences of the source speech and the target speech are aligned by dynamic time warping (DTW), and the alignment information is also used to get the paired one-hot phoneme label. Then feed the label to the adapted average model to generate the paired converted MCEPs. For the training of the error reduction network, the input is the paired converted MCEPs, and the output is the original feature of the target speech. The error reduction network can reduce the error that created by the previous training part.

\subsection{Run-time Conversion Phase}
In the conversion stage, the input is a whole sentence of the source speaker. The conversion of $\log(F_{0})$ and aperiodicity is the same as that of the DBLSTM-based system mentioned in Section II. MFCC features of the source speech are used by the trained ASR model to obtain the one-hot phoneme labels. Next, the one-hot phoneme labels are converted to MCEPs by the trained adapted average model. Then, the converted MCEPs are fed into the error reduction network to get the final result. Finally, the converted MCEPs together with the converted $\log(F_{0})$ and aperiodicity are used by the STRAIGHT vocoder to synthesize the output speech.

\section{Experiments}

\subsection{Experimental Setup}
We conduct listening experiments to assess the performance of our proposed framework that is error reduction network for adapted DBLSTM-based voice conversion. We compare this framework with the baseline DBLSTM \cite{Sun2015} that is explained in Section II, and DBLSTM-based adapted average model that is explained in Section III and given in Fig. 3. We note that the adapted average model is an intermediate step of our proposed algorithm. Fig. 4 also shows the differences between of our proposed framework and the adapted average model. 

The database used in the experiments is CMU ARCTIC corpus \cite{kominek2004cmu}. As it is the most challenging work in voice conversion, we conduct the cross-gender voice conversion experiments. The speech signals are sampled at 16kHz with mono channel, windowed by 25ms, and the frameshift is 5ms. For the DBLSTM-based average model training, data from four male speakers (awb, jmk, ksp, rms) are used. 4433 and 489 sentences are used as training data and validation data. In training phase 2, 45 and 5 sentences from the target speaker (slt) are used as training data and validation data to adapt the average model. For the error reduction training, the same sentences of the target speaker that have been used to adapt the average model and the parallel data of the source speaker (bdl) are used. A DNN-HMM based ASR system \cite{Povey} is used to get the 171-dimension one-hot phoneme label. 40-dimension MCEPs are extracted by STRAIGHT to train the model.

In our proposed approach, to train a DBLSTM-based model, we prefer to use four hidden layers, the number of units in each layer is [171 128 256 256 128 40] respectively. Each bidirectional LSTM hidden layer contains one forward LSTM layer and one backward LSTM layer. The training samples are normalized to zero mean and unit variance for each dimension before training. For the error reduction network, in order to take advantage of context information, three frames of converted MCEPs i.e. current frame, one left frame and 1 right frame are used as input features. In addition, there are three hidden layers in the error reduction network, the number of units in each layer is [120 128 256 128 40] respectively. 

\begin{figure}[t]
	\begin{center}
		\includegraphics[width=87mm]{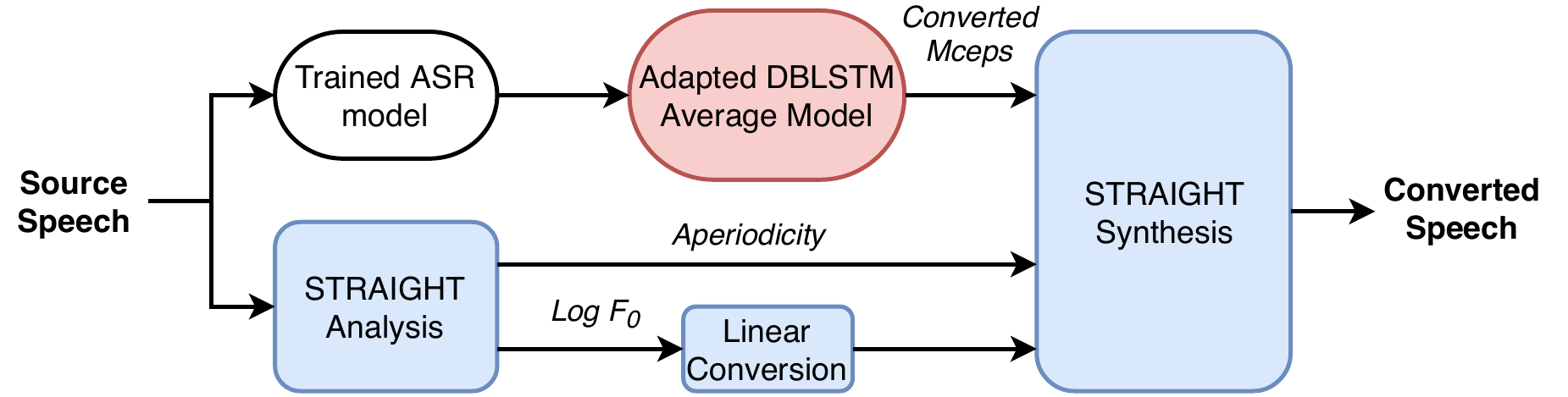}
	\end{center}
	\caption{The run-time phase of the adapted average model.}
	\vspace*{7pt}
\end{figure}

In order to evaluate our proposed approach, 100 parallel utterances from the source speaker and the target speaker are used to train a DBLSTM-based parallel voice conversion system. This system is developed as the baseline approach. There are four hidden layers in the baseline model where the number of units in each layer is same as the training of the adapted average model that is [40 128 256 256 128 40] respectively. We use an open-source CUDA recurrent neural network toolkit CURRENNT \cite{weninger2015introducing} to train the DBLSTM model with a learning rate of $10^{-5}$ and a momentum of $0.9$.

\subsection{Objective Evaluation}

Mel-cepstral distortion (MCD) \cite{mcd1} is used as objective measure of the spectral distance from converted to target speech, which is denoted as:
\begin{equation}
MCD[dB]=\frac{10}{\ln{10}}\sqrt{2\sum_{d=1}^{D}(C_{d}^{target}-C_{d}^{converted})^{2}}
\end{equation}
where $C_{d}^{target}$ and $C_{d}^{converted}$ are the $d^{th}$ dimension of the original target MCEPs and the converted MCEPs, respectively. We expect a good system to report a low MCD value.

The MCD scores of the different systems for the cross-gender voice conversion are summarized in Table I. We can see that our proposed approach outperforms the baseline model and the adapted average model. We can also note that the MCD scores of the adapted average model is not as good as the baseline model, because there is no parallel data in the training of the adapted average model. But after the error reduction network with only 50 utterances of parallel data, the performance can be improved obviously, and outperform both adapted average model and baseline model with 100 utterances of parallel training data.

\begin{table}[h]
	\begin{center}
		\caption{The MCD of different systems.}
		\resizebox{\columnwidth}{!}{%
			\begin{tabular}{ c | c | c | c } 		
				\hline
				\rule[-4pt]{0pt}{14pt}Source-Target & Baseline & Adapted Average Model & Proposed Approach\\
				\hline
				\rule[-4pt]{0pt}{14pt}9.3197 & 6.3042 & 6.7378 & \textbf{6.1989} \\
				\hline
			\end{tabular}
		}
	\end{center}
\end{table}

\subsection{Subjective Evaluation}
To evaluate the quality and similarity of the converted speech from the different systems, we conduct a subjective listening test and 10 listeners are invited to evaluate 10 sentences in each system.

We carry out Mean Opinion Score (MOS) test for evaluating speech quality and naturalness. In the MOS test, comparing with target speech, the grades of the converted speech are: 5 = excellent, 4 = good, 3 = fair, 2 = poor, and 1 = bad. The listeners are asked to rate the speech according to this regulation. In this experiment, we conduct the MOS test among three systems: 1) baseline approach, the parallel DBLSTM-based voice conversion training with 100 utterances of parallel data; 2) adapted average model that explained in Section III; 3) our proposed approach. The results of the MOS test and the 95\% confidence intervals are shown in Fig. 5. The scores of the baseline, adapted average Model and the proposed approach are 2.62, 2.71 and 3.41 respectively.

\begin{figure}[t]
	\begin{center}
		\includegraphics[width=90mm]{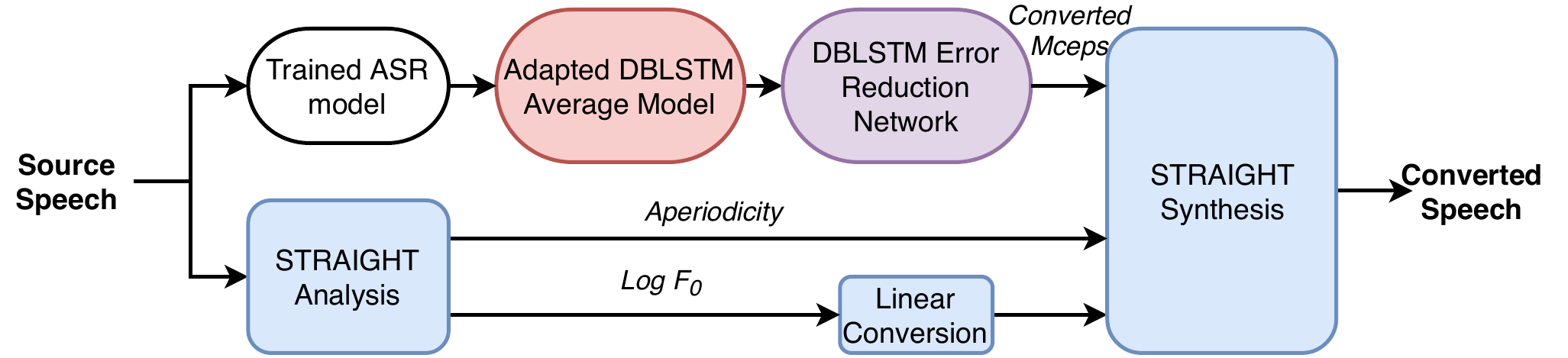}
	\end{center}
	\caption{The run-time phase of the proposed framework.}
	\vspace*{7pt}
\end{figure}

ABX preference test is adopted to evaluate speaker similarity of the converted speech generated by two different systems. The listeners are asked to choose either A or B that sounds closer to the target speaker's speech X. We conduct the ABX preference test between the baseline approach and our proposed approach. The preference bars for speaker similarity are shown in Fig. 6.

\begin{figure}[h]
	\begin{center}
		\includegraphics[width=80mm]{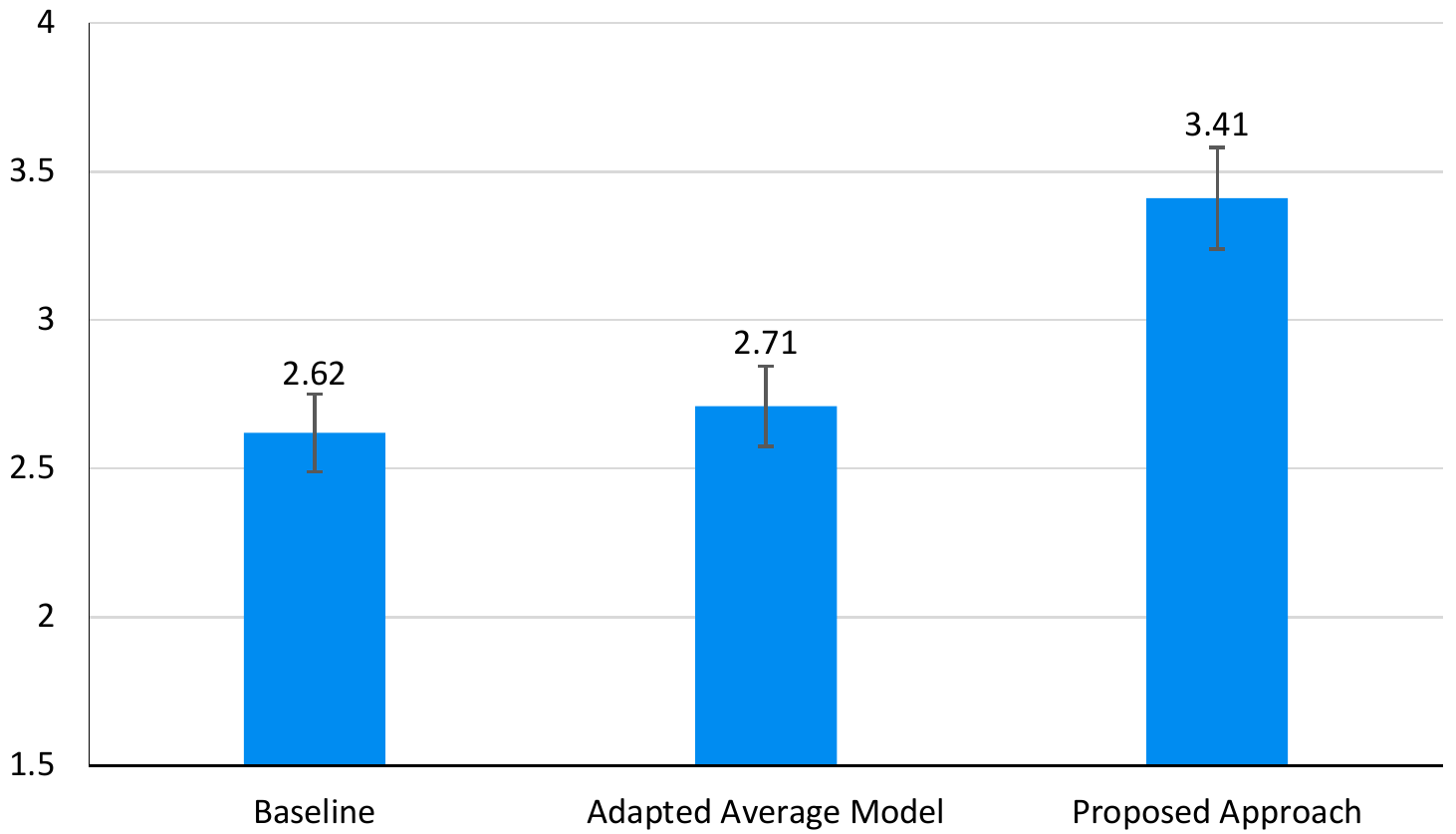}
	\end{center}
	\caption{The result of the MOS test with the 95\% confidence intervals for speech quality and naturalness among the three systems.}
	\vspace*{10pt}
\end{figure}

\begin{figure}[h]
	\begin{center}
		\includegraphics[width=80mm]{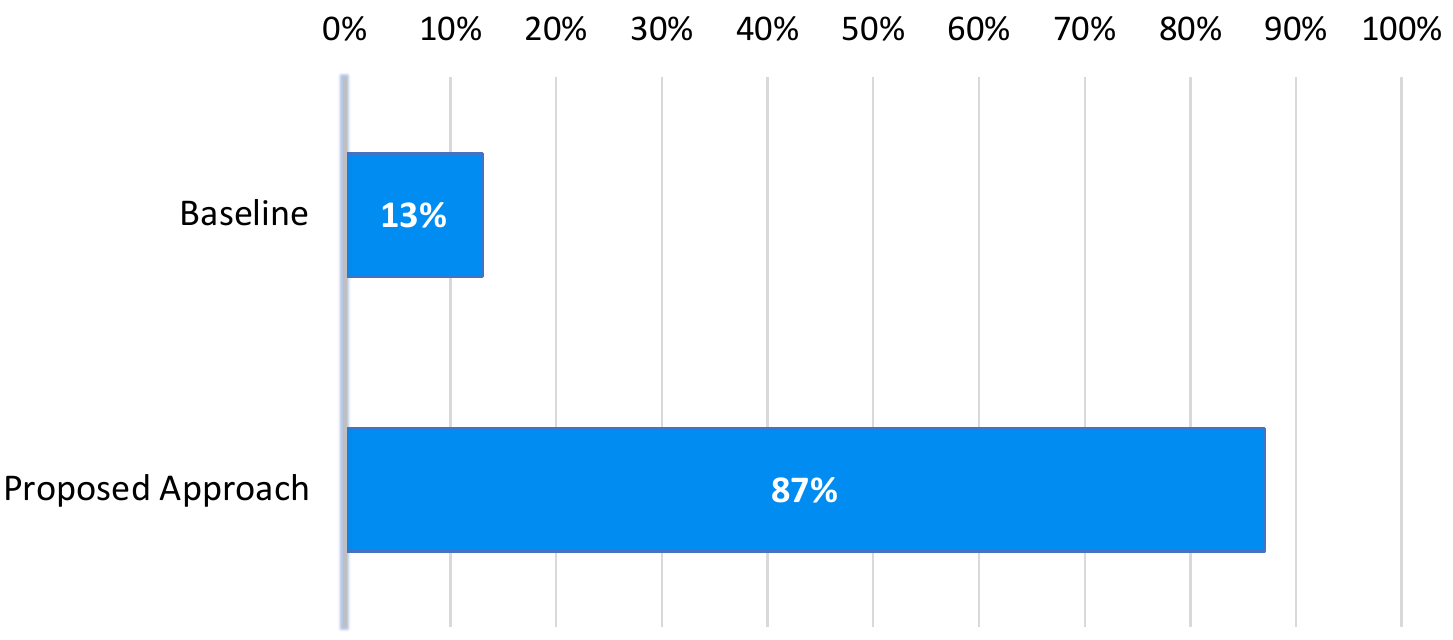}
	\end{center}
	\caption{The result of the ABX preference test for speaker similarity between the baseline and our proposed approach.}
	\vspace*{10pt}
\end{figure}

Overall, the results of both MOS test and ABX preference test show that our proposed error reduction network for adapted DBLSTM-based voice conversion with a limited amount of parallel data outperforms the baseline approach with a large amount of parallel data in both speech quality and similarity. Probably because the average model is trained with a large amount of data to achieve a better speech quality than the baseline approach, improve the performance of the following portions of the system.

\section{Conclusions}
This paper presents an error reduction network for adapted DBLSTM-based voice conversion approach, which can achieve a good performance with limited parallel data of the source speaker and the target speaker. Firstly, we propose to train an average model for the one-hot phoneme label to MCEPs mapping with data from many speakers exclude the source speaker and the target speaker. Then, we propose to adapt the average model with a limited amount of target data. Furthermore, we implement an error reduction network that can improve the voice conversion quality. Experiment results from both objective and subjective evolution show that our proposed approach can make a good use of limited data, and outperforms the baseline approach. In the future, we will investigate to use the WaveNet Vocoder, which is a convolutional neural network that can generate raw audio waveform sample by sample, to improve the quality and naturalness of the converted speech. Some samples for the listening test are available through this link: \url{https://arkhamimp.github.io/ErrorReductionNetwork/}

\section*{Acknowledgment}

This research is supported by Ministry of Education, Singapore AcRF Tier 1 NUS Start-up Grant FY2016. Mingyang Zhang is also supported by the China Scholarship Council (Grant No.201706090063). Berrak Sisman is also funded by SINGA Scholarship under A*STAR Graduate Academy.

\bibliographystyle{IEEEtr}
\bibliography{berrak2}

\end{document}